\documentclass{elsart3}
\usepackage{graphicx}
\usepackage{amssymb}
\usepackage{amsmath}
\newcommand{\be}{\begin{equation}}
\newcommand{\ee}{\end{equation}}
\newcommand{\mibp}{\boldsymbol{p}}
\newcommand{\mibm}{\boldsymbol{m}}
\newcommand{\mibe}{\boldsymbol{e}}
\newcommand{\mibH}{\boldsymbol{H}}
\newcommand{\mibM}{\boldsymbol{M}}
\newcommand{\ext}{\mathrm{ext}}
\newcommand{\eff}{\mathrm{eff}}
\newcommand{\Oe}{\mathrm{Oe}}
\newcommand{\cm}{\mathrm{cm}}

\begin{document}

\begin{frontmatter}



\title{Relaxing-Precessional Magnetization Switching}


\author{Hirofumi Morise and Shiho Nakamura}

\address{Corporate R\&D Center Toshiba Corporation}

\begin{abstract}
A new way of magnetization switching employing both 
the spin-transfer torque and the torque by a magnetic field
is proposed.
The solution of the Landau-Lifshitz-Gilbert equation 
shows that the dynamics of the magnetization in the initial stage of the switching is similar to 
that in the precessional switching,
while that in the final stage is rather similar to the relaxing 
switching.
We call the present method the relaxing-precessional switching.
It offers a faster and lower-power-consuming way of switching 
than the relaxing switching and
a more controllable way than the precessional switching. 
\end{abstract}

\begin{keyword}
spin-transfer torque 
\sep magnetization switching
\sep nanomagnet 
\sep Landau-Lifshitz-Gilbert equation
\sep macrospin model
\PACS 
75.60.Jk
\sep 85.75.-d
\sep 75.75.+a
\end{keyword}
\end{frontmatter}

\section{Introduction}
The high-speed and low-power-consuming memories and data storage devices
are expected in the current days.
One of key technologies realizing such devices is 
the manipulation of the magnetization of a nanomagnet by use of 
the spin-transfer torque, which is proposed by 
Slonczewski \cite{Slonczewski1996} and Berger \cite{Berger1996}.
There, a spin-polarized current is introduced into an assembly of nanomagnets.
The spin-polarized current, whose polarization direction is $\hat\mibp$,
acts on a nanomagnet whose magnetization direction vector is $\hat\mibm$.
The torque caused by the current is proportional to the vector 
$\hat\mibm\times(\hat\mibp \times\hat\mibm)$ \cite{Slonczewski1996}.
It fluctuates the magnetization \cite{Tsoi1998} 
or even reverse its direction \cite{Myers1999}.
\par
To date, two different methods of the magnetization switching
utilizing the spin-transfer torque have been proposed.
The first method, which we call the relaxing switching,
is more common.
There, the spin-polarization of the current which acts on the 
nanomagnet is almost parallel (or antiparallel) to
the magnetization.
The magnetization switches its direction according to the directions of 
the spin-polarization and the flow of the electrons.
The switching dynamics is understood as a result of a relaxing process toward the equilibrium state.
The main problem in the method for the application to memories is
that it requires relatively high current density.
The critical dc current, $J_c$, is of the order of $10^7 \mathrm{A}/\cm^2$.
A naive explanation for 
the need of high current density is the following.
The magnitude of the torque is small
when the relative angle $\theta$ between the magnetization of the nanomagnet 
and the spin-polarization of the conduction electrons is small,
since $|\hat\mibm\times(\hat\mibp \times\hat\mibm)|=\sin\theta \approx \theta 
\ll 1$.
\par
Another problem is that the switching is slow.
The results of the simulation indicates that the magnetization repeats the precessional 
motion until it relaxes to the equilibrium state, demanding a
time of the order of 1 ns in the case where the current is 
$2$-$3 J_c$ \cite{Sun2000,Li2003}.
An experimental result show that at least four times larger current than the 
calculated critical current is needed to
switch the magnetization in 100 ps \cite{Tulapurkar2004}.
In summary, the relaxing switching
does not bring out the full potential ability of
the spin-transfer torque.
\par
The second method, which are called the precessional switching,
has a possibility of overcoming such disadvantages \cite{Kent2004,LeeKJ2005}. 
There, the spin-polarization of the current is perpendicular to the 
magnetization of the nanomagnet.
By taking such a configuration, it is possible to transmit the spin-transfer torque 
to the nanomagnet efficiently, and therefore realize the fast and 
low-power-consuming magnetization switching.
However,
it requires high-level controllability of the shape of the current pulse, including the 
pulse amplitude and the temporal width.
Furthermore,
the final state of the magnetization depends on its initial 
state, which indicates that the read-before-write is necessary.
These disadvantages are due to the fact that the precessional switching employs
the nonequilibrium state of the magnetization rather than the
relaxation to the equilibrium state.
\par
In the present work, an alternative method, which resolves the problems of the 
previous methods, is proposed.
We call this new method the relaxing-precessional switching.
It is different from the previous methods since it requires the simultaneous
introduction of the spin-polarized current and the external magnetic field.
The stable magnetization states under the spin-polarized current and the 
external magnetic field have been
exhaustively studied in the previous work \cite{Morise2005}.

\section{Phase diagram based on the linear stability}
\label{sec:linear}
We consider a system composed of two ferromagnetic layers separated by a 
nonmagnetic layer.
When the current in the direction perpendicular to the plane is introduced,
the magnetization of the thinner (free) layer receives torques from the current,
which is spin-polarized in the direction $\hat\mibp$. 
Here, $\hat\mibp$ denotes the magnetization direction vector of 
the thicker (pinned) layer.
In addition, an external magnetic field $\mibH_\ext$ is applied in the in-plane hard axis direction.
The magnetization direction vector $\hat\mibm=\mibM/M$ of the free layer obeys the following LLG equation:
\begin{align}
\frac{d\hat\mibm}{dt}=&
-\alpha\hat\mibm \times \frac{d\hat\mibm}{dt}
+\gamma\hat\mibm \times \mibH_\eff
\nonumber\\
&+2\pi M\gamma j \hat\mibm\times(\hat\mibp \times \hat\mibm),
\label{LLG}
\end{align}
where
\begin{align}
&\mibH_\eff
\equiv  \mibH_\ext-4\pi M m_x \mibe_x +H_Km_z\mibe_z,
\nonumber\\
&j=\frac{1}{2\pi M \gamma}\frac{2\mu_B}{VM}\frac{gI_e}{|e|}.
\end{align}
\par
\begin{figure}
\begin{center}
\includegraphics[scale=0.7]{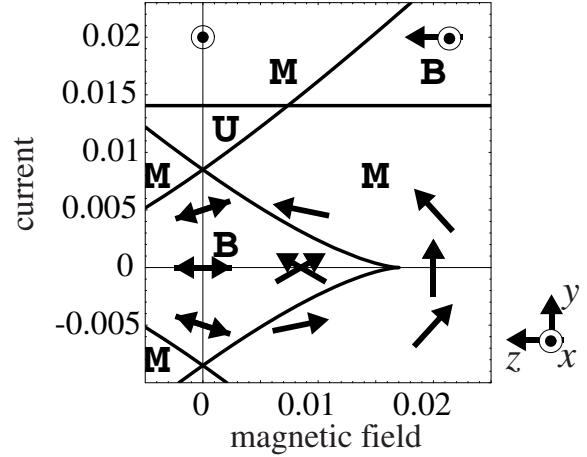}
\end{center}
\caption{
 Phase diagram of stable states based on the linear analysis.
 The horizontal and the vertical axes denote
 the dimensionless magnetic field $H_\ext/2\pi M$
 and the dimensionless current $j$, respectively.
 We set $H_K/2\pi M=150\Oe/0.9T=0.017$.
 Each of the arrows and the circles with dots 
 represents the direction of a stable magnetization state.
 The former (latter) corresponds to the direction in (out of) 
 the $yz$-plane.
 The labels ``B'', ``M'', and ``U'' denote bistable, monostable, 
 and unstable regions, respectively.
}
\label{pd-sta}
\end{figure}

Here, $H_K$, $\alpha$, $\gamma$, $\mu_B$, $V$, $I_e$, and $g$ denote
the anisotropy field, the damping constant, the gyromagnetic ratio, 
the Bohr magneton,
the volume of the free layer, the electric current, 
and the efficiency factor\cite{Slonczewski1996},
respectively.
The directions of $\mibH_\ext$ and $\hat\mibp$ are chosen so that 
$\mibH_\ext=H_\ext \hat\mibe_y$ and $\hat\mibp=\hat\mibe_x$
respectively
in the relaxing-precessional switching.
The linear stability condition for the equilibrium states of the 
magnetization \cite{Morise2005}
allows one to draw the phase diagram for the stable magnetization states, 
which is shown in Fig. \ref{pd-sta}.
We notice that there are bistable (B), monostable (M), and unstable (U) regions.
In particular, we pay a special attention to
the transition 
from a bistable region to a monostable region
since it can be used as a switching mechanism.
When the system undergoes such a transition by introducing 
the magnetic field and the spin-polarized current,
the final direction of the magnetization is uniquely determined
regardless of the initial direction.
This gives the relaxing-precessional switching an advantage
over the precessional switching.
In addition,
Fig.\ref{pd-sta} indicates that the direction of the magnetization
can be changed to desirable direction between the two 
by choosing the direction of the spin-polarized current.
\par
The critical current $j_c^\mathrm{RP}$, needed for the reversal, 
is given by the boundary line
between the bistable region and the monostable region.
The monostable region appears
because an equilibrium state, which was labeled as (S) 
in Ref.\cite{Morise2005}, become unstable
above the boundary line.
The analytical expression for this line is obtained by solving the equation
(see Appendix for the derivation):
\begin{align}
 H_\ext=H_K\cdot h(j_{c}^\mathrm{RP}\cdot 2\pi M / H_K),
 \label{jc-rp}
\end{align}
where $h(\xi)$ is a function defined by 
\begin{align}
h(\xi)\equiv\xi X(\xi)^{-3/2}(1-2X(\xi)),
\end{align}
and $X(\xi)$ is a root of the equation : $X^3+\xi^2 (X-1)=0$.
The asymptotic form of $j_c^\mathrm{RP}$ around $H_\ext=0$ and that 
around $H_\ext=H_K$ are useful. They are written as
\begin{align}
& j_c^\mathrm{RP}(H_\ext \approx 0)
= \frac{H_K}{2\pi M}
\times\nonumber\\
& \left\{\frac{1}{2}-\frac{1}{\sqrt{2}}\frac{H_\ext}{H_K}
 +\frac{1}{8}\left(\frac{H_\ext}{H_K}\right)^2+O\left(\frac{H_\ext}{H_K}\right)^3\right\}
 \label{jc_near_zero_field}
\end{align}
and
\begin{align}
& j_c^\mathrm{RP}(H_\ext \approx H_K)=\frac{H_K}{2\pi M}
\left[\frac{2}{3}\left(1-\frac{H_\ext}{H_K}\right)\right]^{3/2}
\times\nonumber\\
&
\left\{1-\frac{1}{12}\left(1-\frac{H_\ext}{H_K}\right)
+\frac{1}{864}\left(1-\frac{H_\ext}{H_K}\right)^2\right.
\nonumber\\
&\left.+O\left(1-\frac{H_\ext}{H_K}\right)^3\right\},
\end{align}
respectively.
\par
We notice that
the critical current for the precessional switching
\cite{LeeKJ2005,Morise2005}
 $j_c^\mathrm{P}=H_K/4\pi M$ coincides with 
$j_c^{RP}(H_\ext=0)$ by setting $H_\ext=0$ in eq.(\ref{jc_near_zero_field}).
As the field $H_\mathrm{ext}$ increases from zero, 
the critical current $j_c^\mathrm{RP}(H_\ext)$
decreases from this value. 
Taking $H_\ext=0.5 H_K$, for example,
we obtain $j_c^\mathrm{RP}=0.37j_c^\mathrm{P}$.
Furthermore, by taking 
$\alpha=0.01$ and
$H_K/4\pi M=150 \Oe/1.8T=0.0083$
as typical values,
the critical currents of the relaxing-precessional,
relaxing, and precessional switching are
$j_c^\mathrm{RP}=0.003,\enskip
j_c^\mathrm{R}(=\alpha)=0.01,\enskip
j_c^\mathrm{P}=0.008,
$
respectively, which implies that
the relaxing-precessional switching
is the lowest-power consuming way among the three methods.

\section{Numerical calculation of the LLG equation}
\label{sec:range}
As was discussed in the previous work \cite{Morise2005},
the final state which the magnetization obtains is beyond the framework of 
the linear stability.
Therefore, the numerical calculation of the LLG equation is necessary.
In addition, switching speed is also discussed from the results. 
Here, it is assumed that the magnetic field is applied 
synchronously with the spin-polarized current.
\par
\begin{figure}
 \includegraphics[scale=0.6,bb=140 320 480 530]{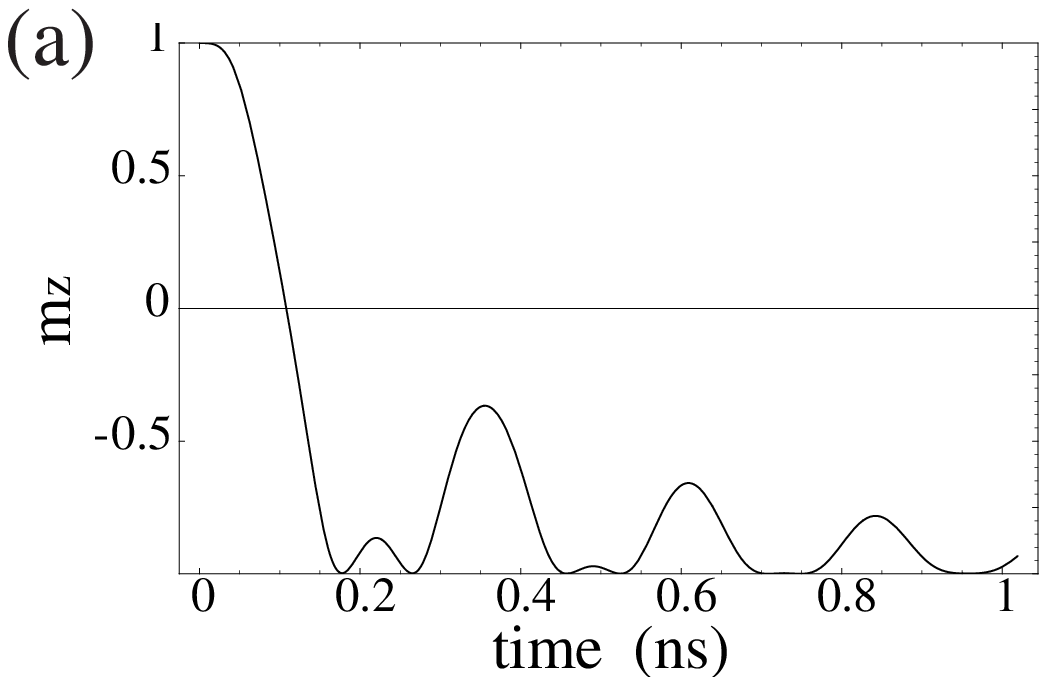}
 \includegraphics[scale=0.6,bb=140 320 480 530]{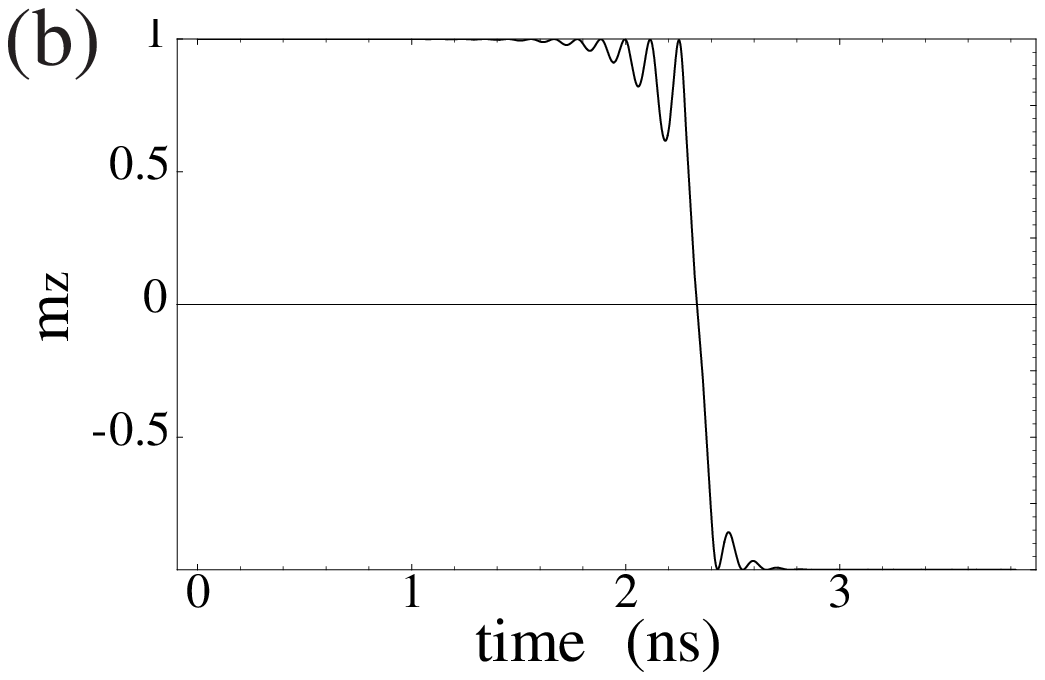}
 \includegraphics[scale=0.6,bb=140 320 480 530]{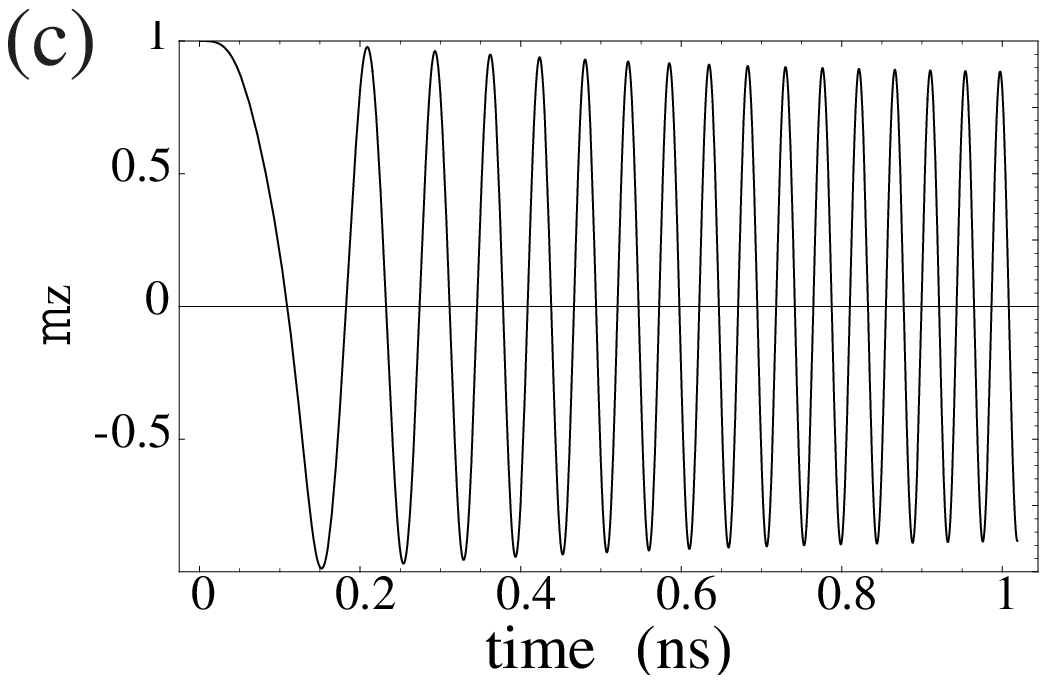}
 \caption{The time evolution of $m_z$ in
 (a) relaxing-precessional switching with $H_\ext=0.5 H_K$ and $j=0.003$, 
 (b) relaxing switching with $j=0.03$, 
 and (c) precessional switching with $j=0.01$.
 We set $H_K/2\pi M=0.017$.
 }
 \label{dyn}
\end{figure}

Fig.\ref{dyn}(a) shows a time-dependent solution of the LLG equation on the basis 
of the macrospin (single domain) model.
In this example, the parameters are taken as $H_\ext=0.5H_K$ and $j=0.003$.
It is worth comparing this solution with (b)that in the relaxing switching or 
(c)that in the precessional switching.
\par
In the initial stage of the relaxing-precessional switching, 
the magnetization dislocates its direction from the initial one
and begins precessional motion in response to the introduction of 
the spin-polarized current.
This dynamic behavior is similar to that in the precessional switching.
In the typical example of the relaxing-precessional switching 
shown in Fig.\ref{dyn}(a),
the reversal time is approximately 0.2 ns,
which is almost the same as that in the precessional switching
with a larger current [see Fig.\ref{dyn}(c)].
\par
Once the magnetization vector
gets close to the opposite direction,
it relaxes to the destination.
The dynamics in this stage is very much different from
that in the precessional switching and rather similar to that in the 
relaxing switching.
This phenomenon is advantageous for the application to memories
since it implies that any strict control
of the temporal width of the current pulse is unnecessary.

\section{Ranges of the current and the magnetic field}
The next task is to reveal the ranges of the spin-polarized current and the magnetic field
which can be used as the relaxing-precessional switching.
Here, we assume that the magnetic field is applied constantly in time.

\begin{figure}
 \includegraphics[scale=0.65,bb=120 550 450 800]{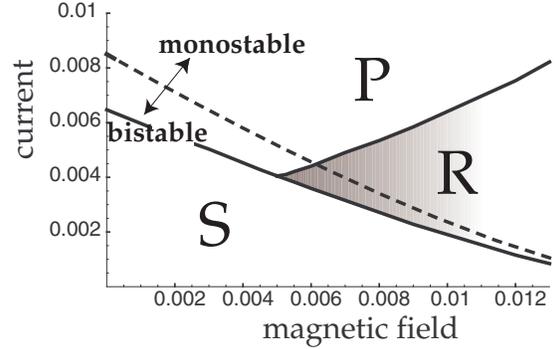}
 \caption{
 Phase diagram of the dynamical behavior of the magnetization.
 We set $H_K/2\pi M=0.017$.
 The definition of the axes is the same as that in Fig.\ref{pd-sta}.
 The solid lines represent the boundary lines 
 which distinguish the final states of the magnetization.
 In ``S'', the magnetization stays in the neighborhood of the initial direction.
 In ``P'', the magnetization continues a precessional motion.
 In ``R'', the magnetization reverses its direction.
 The dashed line represents the boundary line 
 between the monostable region and the bistable one, which have been shown
 in Fig.\ref{pd-sta}.
 }
 \label{pd-dyn}
\end{figure}
\par
Fig.\ref{pd-dyn} is a phase diagram which illustrates the type of
the dynamical behavior of the magnetization
after the long-time introduction of the given magnitude of the 
spin-polarized current and that of the magnetic field.
We have obtained this phase diagram by performing
numerical calculations of the macrospin LLG equation.
In the simulation, the current is assumed to
rise instantaneously and be kept a constant value since then.
The whole region is divided into one of the three regions 
labeled by ``S'', ``R'', and ``P'', respectively.
In the case where the point corresponding to the given values of the 
spin-polarized current and the magnetic field lies in the region S, 
the magnetization stays in the neighborhood of the 
initial direction
during the introduction of the 
spin-polarized current.
In the case where it lies in the region R, 
the magnetization reverses its direction and does not 
return to the initial one after the spin-polarized current 
is introduced.
In the case where it lies in the region P, 
the magnetization precesses and approaches neither the 
initial direction nor the opposite one.
Therefore, the region R denotes the ranges of the spin-polarized current
and the magnetic field which can be used 
as the relaxing-precessional switching.
The dashed line represents the boundary between the monostable and the 
bistable states, which have been shown in Fig.\ref{pd-sta}. 
\par
We have two remarks.
First,
the monostable region in Fig.\ref{pd-sta} 
does not overlap with the region S in Fig.\ref{pd-dyn}.
This implies that the monostability is a sufficient condition of the switching.
Thus, the analytical expression of the critical current obtained in Sec.\ref{sec:linear}
overestimates the actual switching current.
\par
Second, the ranges of the spin-polarized current and the magnetic field
for the relaxing-precessional switching are limited to
the triangle-shaped region shown as darkened one in Fig.\ref{pd-dyn}.
When the spin-transfer torque is large,
the switching is realized as a precessional switching rather than 
a relaxing-precessional switching.
Therefore, too large spin-polarized currents are not suitable.
Too large magnetic field is also undesirable for practical use.
The magnetic field can be applied either 
by a current field or by a permanent magnet.
In the former case, an additional power-consumption
besides the current flowing through the multilayer
must be taken into account while
no additional power is consumed in the latter case.
\par
To design a practical device including the read-out function, 
we can add a magnetic reference layer,
whose magnetization direction is collinear to that of the free layer,
to the system.
In this system, the effect of the reference layer 
to the spin-transfer torque is negligible since
the magnitude of the current used in the relaxing-precessional switching
is too small to cause the relaxing switching as explained 
in the previous sections.
Therefore, the adequate ranges of the magnetic field and the spin-polarized 
current discussed in this section are not affected by the addition of the reference layer.

\section{Summary}
In summary, we have proposed the relaxing-precessional magnetization switching,
which utilizes both the spin-transfer torque and torque by a magnetic field.
The present method 
can reverse a magnetization 
one order of magnitude faster with a current one order of magnitude lower
than the conventional relaxing switching.
In addition, it does not require
strict control of the current pulse 
unlike the precessional switching.

\appendix

\section{Derivation of eq.(\ref{jc-rp})}
\label{app}
In this appendix, we present the derivation of 
the analytical expression (\ref{jc-rp}) of the boundary line
between the region M and B.
We concentrate on the case where $j>0$ and $0<H_\ext < 
H_K \ll 2\pi M$. 
First, we parameterize the vector $\mibm$, 
which is on a unit sphere, in terms of
the spherical coordinates 
$\theta\in[0,\pi]$ and $\phi\in[0,2\pi)$ as
\begin{align}
m_x=  \sin\theta\cos\phi,\enskip
m_y= \cos\theta,\enskip
m_z=-  \sin\theta\sin\phi.
\end{align}
The LLG equation (\ref{LLG}) is rewritten as
\begin{align}
 \begin{pmatrix} \dot{\theta} \\ \dot{\phi} \end{pmatrix} = &
 \begin{pmatrix} f(\theta,\phi) \\ g(\theta,\phi) \end{pmatrix}
 \equiv
 \frac{\gamma}{1+\alpha^2}\begin{pmatrix} 1 & \alpha \\ -\alpha & 1 \end{pmatrix}
\begin{pmatrix}f_1 \\ f_2 \end{pmatrix}
\nonumber\\
f_1=&
   \cos\phi(
   - 4\pi M \sin\theta\sin\phi
   - H_K \sin\theta\sin\phi
   \nonumber\\ 
 &+ 2\pi M j \cos\theta ) 
   \nonumber\\ 
f_2=&
   (H_K \sin^2\phi- 4\pi M \cos^2\phi)
   \sin\theta\cos\theta
\nonumber\\
   &- H_\ext \sin\theta  
   - 2\pi M j\sin\phi
\end{align}
in the spherical coordinate.
The equilibrium states are given by the solutions of the equations:
$f(\theta,\phi)=g(\theta,\phi)=0$.
The solution which becomes unstable for large enough
$j>0$ and $H_\ext$ is labeled as (S) in Ref.\cite{Morise2005},
which
continuously changes into $\mibm=-\mibe_y$
as $H_\ext \to {}^\exists H<-H_K$ and $j\to 0$.
The solution (S) is obtained by
\be
H_K \sin\theta \cos\theta -H_\ext \sin\theta - 2\pi M |j|=0,
\enskip \sin\phi= \mathrm{sgn}(j).
\label{app:solS}
\ee
In the following, we consider the existence of this solution
and the stability of the state corresponding to this solution.
The equation (\ref{app:solS}) is equivalent to
\be
\pm \sqrt{1-x^{-2}}=(2\pi M j/H_K) x+H_\ext/H_K,
\ee
if we  set $x\equiv 1/\sin\theta$.
Thus,
the existence of the solution $\theta$ is guaranteed when
the two lines in the $xy$-plane, $L_1$ : $y=\pm \sqrt{1-x^{-2}}$ and 
$L_2$ : $y=(2\pi M j/H_K) x +H_\ext/H_K$,
share at least one point.
\begin{figure}
\includegraphics[scale=0.65,bb=60 620 380 850]{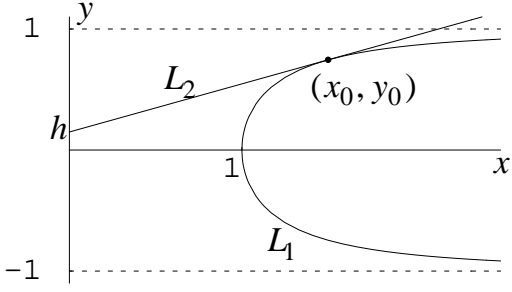}
\caption{$L_1$ : $y=\pm \sqrt{1-x^{-2}}$ and 
$L_2$ : $y=(2\pi M j/H_K) x +H_\ext/H_K$.
When the two lines have common points, the equilibrium states 
corresponding to eq.(\ref{app:solS})} exist.
\label{app:fig:tangent}
\end{figure}
Reminding that the gradient of the line $L_2$ is positive,
we see from Fig.\ref{app:fig:tangent} that
the two lines intersect when
the $y$-intercept of the line $L_2$ is smaller than some critical value 
$h(2\pi M j / H_K)$, which is a function of the gradient of $L_2$.
In other words, the existence condition is written as
\be
\frac{H_\ext}{H_K} \leq h\left(\frac{2\pi M j}{H_K}\right).
\label{app:existence-condition}
\ee 
In the inequality (\ref{app:existence-condition}), 
the equality holds when $L_1$ and $L_2$ are tangent to each other.
In that case, the tangent point $(x_0,y_0)\equiv(1/\sin\theta_0,\cos\theta_0)$ satisfies
\begin{align}
 y_0=\sqrt{1-x_0^{-2}}=&\frac{2\pi M j}{H_K}x_0
 +h\left(\frac{2\pi M j}{H_K}\right),
\label{app:tangent1}
\end{align}
\begin{align}
& \left.\frac{d}{dx}\sqrt{1-x^{-2}}\right|_{x=x_0}
\nonumber\\
&=\left.\frac{d}{dx}\left(\frac{2\pi M j}{H_K}x+
h\left(\frac{2\pi M j}{H_K}\right)\right)\right|_{x=x_0}.
\label{app:tangent2}
\end{align}
The condition (\ref{app:tangent2}) yields
\be
x_0^{-6}+\left(\frac{2\pi M j}{H_K}\right)^2(x_0^{-2}-1)=0,
\ee
which is rewritten as
\be
\frac{2\pi M j}{H_K}=\frac{x_0^{-3}}{\sqrt{1-x_0^{-2}}}=\frac{\sin^3\theta_0}{\cos\theta_0}.
\ee
By introducing the function $X(\xi)$, which is defined as
a root of the equation : $X^3+\xi^2 (X-1)=0$, we have 
$x_0^{-2}=X(2\pi M j/H_K)$ and
\begin{align}
 h\left(\frac{2\pi M j}{H_K}\right)
 =&\frac{2\pi M j}{H_K}\cdot
 X\left(\frac{2\pi M j}{H_K}\right)^{-3/2}\times
\nonumber\\
& \left(1-2X\left(\frac{2\pi M j}{H_K}\right)\right).
 \label{app:hcrit}
\end{align}
When $H_\ext <H_K \cdot h(2\pi M j/H_K)$,
$L_1$ and $L_2$ intersect each other at two points.
Therefore, eq. (\ref{app:solS}) have two solutions $\theta_<$ and $\theta_>$,
which satisfy $0<\theta_<<\theta_0<\theta_><\pi/2$. 
\par
In general, the equilibrium state is stable if all eigenvalues of
the matrix:
 \begin{align}
  L=\left.\begin{pmatrix}
     \partial f /\partial \theta & \partial f / \partial \phi \\ 
     \partial g /\partial \theta & \partial g / \partial \phi 
 \end{pmatrix}\right|_{f=g=0}
 \end{align}
have negative real parts, in other words, $\mathrm{tr} L <0$ and $\det L>0$.
In the present system, the trace of $L$ is negative because 
\begin{align}
\frac{1}{\alpha}\mathrm{tr} L=&
-2+
\frac{H_K}{2\pi M}(3\cos^2\theta-2)
-2\cdot \frac{H_\ext}{2\pi M}\cos\theta
\nonumber\\
\leq & -2+\frac{H_K}{2\pi M}
+2\cdot \frac{H_\ext}{2\pi M}
<0.
\end{align}
Here, we have used the relation (\ref{app:solS}).
\par
In contrast, the sign of the determinant of $L$ cannot be determined automatically.
Since $\det L$ is calculated as
\begin{align}
\det L=&(1+\alpha^2)
\left(2+\frac{H_K}{2\pi M}\sin^2\theta+\frac{H_\ext}{2\pi M}\cos\theta\right)\times\nonumber\\
&\left(\frac{H_K}{2\pi M} \sin^2\theta
-j \frac{\cos\theta}{\sin\theta}\right),
\end{align}
we obtain
\begin{align}
\mathrm{sgn}(\det L)=&\mathrm{sgn}(
\frac{\sin^3\theta}{\cos\theta}-\frac{2\pi M j}{H_K})
\nonumber\\
=&\mathrm{sgn}(\frac{\sin^3\theta}{\cos\theta}-\frac{\sin^3\theta_0}{\cos\theta_0})
=\pm 1 \Longleftrightarrow \theta \gtrless \theta_0.
\end{align}
Thus, the equilibrium state corresponding to the solution $\theta_>$ is 
always stable if it exists. In contrast, that corresponding to $\theta_<$
is always unstable.
\par
In summary, we have proved that there exist two equilibrium states
under the condition $H_\ext \leq H_K\cdot h(2\pi M j/H_K)$.
One of the equilibrium states is always stable
if it exists.
Namely, the above condition is not only 
the existence condition of a equilibrium state 
but also its stability condition.







\end{document}